\documentstyle[graphicx,aps,twocolumn]{revtex}
\newcommand{\bea}{\begin{eqnarray}}
\newcommand{\beq}{\begin{equation}}
\newcommand{\eea}{\end{eqnarray}}
\newcommand{\eeq}{\end{equation}}
\begin{document}
\title{Dynamical moment of inertia and quadrupole vibrations 
in  rotating nuclei}
\author{\sc R.G. Nazmitdinov$^{1,2}$ 
D. Almehed $^{3}$, F.~D\"onau $^4$}
\address{\sl 
$^1$  Max-Planck-Institut f\"ur Physik Komplexer Systeme, D-01187 Dresden, 
Germany\\ 
$^2$ Bogoliubov Laboratory of Theoretical Physics
Joint Institute for Nuclear Research, 141980 Dubna, Russia\\
$^3$ Department of Physics, UMIST, P.O. Box 88,
Manchester M60 1QD, United Kingdom\\
$^4$Institut f\"ur Kern- und Hadronenphysik, 
FZ Rossendorf, 01314 Dresden, Germany}
\maketitle

\vspace{0.3cm}

\begin{abstract}
{\sf The contribution of quantum shape fluctuations 
to inertial properties of rotating nuclei has been analysed 
within the self-consistent one-dimensional cranking oscillator
model. It is shown that
in even-even nuclei the dynamical moment of inertia calculated in 
the mean field approximation is equivalent to the Thouless-Valatin moment
of inertia calculated in the random phase approximation 
{\it if and only if} the self-consistent conditions for the mean field
are fulfilled.
}   
\end{abstract}

\vspace{0.2in}


\vspace{0.2in}

A description of rotational states 
is  one of the oldest, yet not fully solved, problem in 
nuclear structure physics.
While various microscopic models based on the 
cranking approach \cite{RS,Fr01} describe reasonably well 
the kinematical moment of inertia 
${\cal J}^{(1)} = -(dE/d\Omega)/\Omega $ 
for a finite angular frequency $\Omega$,  
there is  still a systematic deviation of the dynamical moment of inertia
${\cal J}^{(2)}=-d^2E/d\Omega^2$ (E is the energy in the rotating frame)
from the experimental data at high spins \cite{Af}. 
Since the moments of inertia are the  benchmarks 
for microscopic models of collective motion in nuclei, 
the understanding of the source of the discrepancy becomes 
a challenge for a many body theory of finite Fermi systems.
 
The pairing correlations introduced in nuclear
physics by Bogoliubov \cite{Bog} and Bohr, Mottelson and 
Pines \cite{BMP} improved the description
of the inertial nuclear properties,  especially 
in the low spin region \cite{Bel}. 
It was conjectured that at high spins
the Coriolis and centrifugal forces break
Coopers pairs from the pairing condensate  and cause
the transition from  the superfluid 
to the normal (unpaired) fluid \cite{MV}. 
As a consequence, the rigid body (kinematical) 
moment of inertia should be reached 
at the normal phase.  However, even for superdeformed nuclei where
the pairing condensate is expected to be strongly quenched, 
the moment of inertia is usually lower than the rigid body 
value.

In fact, the conjecture was based on  
the analogy between moment of inertia of a rotating nucleus 
and  magnetic susceptibility of a macroscopic superconductor
under a magnetic field. Therefore, the conjecture may lose its validity 
for finite Fermi systems like nuclei, while remaining 
correct for an infinite number of particles.
For example, the deviation from the rigid body value 
could be partially explained due to shell effects \cite{Fr}. 
In present paper we demonstrate that correlations caused
by shape oscillations of a system are another important
ingredient which is missing in all state of the art
calculations of the moments of inertia for rotating nuclei.
We focus our analysis upon the dynamical moment of inertia, since
the ${\cal J}^{(2)}$
contains more information about different properties of the system
due to the obvious relation
${\cal J}^{(2)}={\cal J}^{(1)}+
\Omega d{\cal J}^{(1)}/d\Omega$.

The first attempt to take into account the contribution of the quadrupole 
oscillations to the correlation energy and to the moment of inertia
within the cranking+random phase approximation (RPA) approach  \cite{Eg} 
suffered from an inconsistency between
the mean field and the residual interaction. In addition, 
the calculations were done in a restricted configuration
space (only 3 shells have been considered). 
For a time the solution of the problem was postponed, 
since there was no practical
recipe to calculate the contribution of the correlation energy, 
even in a restricted configuration space. 
Using the integral representation method developed recently
in \cite{dan}, a full energy $E_{RPA}$ can be calculated in the 
RPA order \cite{RS,BR} with a high accuracy and with  
minimal numerical efforts. Consequently, this energy can be used
to calculate the dynamical moments of inertia  ${\cal J}^{(2)}$. 
On the other hand, in literature it is stated
that the dynamical moment of inertia should be equivalent to
the Thouless-Valatin moment of inertia \cite{Th}.
Since the realistic application of the Thouless-Valatin theory requires the
self-consistent solution of the mean field  and the RPA equations, 
until now this point is not clarified. 
 
To understand the role of shape oscillations upon the value of the moment of
inertia and to calculate the Thouless-Valatin moment of inertia in a simple
but still realistic model,  we use a self-consistent cranked
triaxial harmonic oscillator model as a rotating mean field.
The self-consistent residual interaction constructed 
according to the recipe \cite{Ab} is added to describe the
collective excitations in a rotating system. 
Since  all shells are mixed, we go beyond the
approximation used in \cite{Eg} 
(for a cranking harmonic oscillator see also \cite{Nil}). 
Notice that the contribution of 
the pairing vibrations to the correlation energy aside 
of the one from the shape vibrations  is  also important 
(see \cite{A0,Ber} and references there). However, there are some open
problems with the choice of  the self-consistent pairing interaction.
Therefore, the combined effect
of the both types of vibrations is beyond the scope of 
the present investigation
and we leave this problem for the future.

The mean field part of the many-body Hamiltonian (Routhian) 
in the rotating frame is given by
\beq
\label{mf}
H=\sum_{i=1}^N(h_0-\Omega l_x)_i
=H_0-\Omega L_x
\eeq
where the single-particle triaxial harmonic oscillator Hamiltonian $h_0$ is
aligned along its principal axes and reads
\beq
h_0={1\over 2m}{\vec  p}^2+{m\over 2}
(\omega _x^2 x^2+\omega _y^2 y^2+\omega _z^2 z^2).
\eeq
The eigenmodes and the total energy of 
the mean field Hamiltonian Eq.(\ref{mf})
are well known \cite{Val,RBK,Zel}, 
\beq
\label{fr}
\omega_{\pm}^2=\frac{1}{2}\left(\omega_y^2+\omega_z^2+2\Omega^2\pm 
[(\omega_y^2-\omega_z^2)^2 
+8\Omega^2(\omega_y^2+\omega_z^2)]^{1/2}\right)
\eeq
\beq
\label{en}
E_{MF}=\hbar\left(\omega_x{\sum}_x+\omega_+{\sum}_+ 
+ \omega_-{\sum}_-\right).
\eeq
Here, 
$\sum_k=\langle \sum_j^N(n_k+1/2)_j \rangle$ and $n_k=a_k^+a_k$ ($k=x,+,-$)
where $a_k^+$, $a_k$ are the oscillator quanta operators.  
The lowest levels are filled from the bottom,
which give the ground state energy in the rotating frame.
The Pauli principle is taken into account such that
two particles occupy one level.
The minimisation of the total energy Eq.(\ref{en})
with respect to all three frequencies,
subject to the volume conservation condition 
$\omega_x\omega_y\omega_z=\omega_0^3$, yields the
self-consistent condition \cite{TA,HN}
for a finite rotational frequency
\beq
\omega_x^2\langle x^2 \rangle =\omega_y^2 \langle y^2 \rangle =\omega_z^2
\langle z^2 \rangle .
\label{cond}
\eeq
It should be pointed out that the condition Eq.(\ref{cond}) provides
generally the absolute minima in comparison with the local minima obtained
from the condition of the {\it isotropic velocity distribution} \cite{RBK,Zel} 
\beq
\label{con1}
{\sum}_x\omega_x={\sum}_+\omega_+={\sum}_-\omega_-
\eeq
at large rotational frequency. 
 
To analyse the contribution of the quadrupole 
shape oscillations we add to the mean 
field Hamiltonian Eq.(\ref{mf}) the self-consistent 
interaction resulting from small angular rotations
around the $x-$, $y-$, $z-$ axes and small variations 
of two the intrinsic shape parameters $\varepsilon$ and 
$\gamma$ \cite{Ab}.
Consequently, the total Hamiltonian can be expressed as
\beq
H_{{\rm RPA}}=H_0-\Omega L_x
- {\kappa \over 2}\sum _{\mu =-2}^2 Q_{\mu }^{\dagger} Q_{\mu }
={\tilde H} - \Omega L_x.
\label{hrpa}
\eeq
The effective interaction restores the rotational
invariance of the Hamiltonian $H_0$ such  that now
$[{\tilde H},L_i]=0 \quad  (i=x,y,z)$ in the RPA order.
The self-consistency condition
Eq.(\ref{cond}) fixes the quadrupole strength
$\kappa=\frac{4\pi}{5}\frac{m\omega_0^2}{\langle r^2 \rangle}$.
Here a mean value
$\langle r^2 \rangle = \langle \bar x^2+\bar y^2+\bar z^2 \rangle$
and quadrupole operators $Q_{\mu}={\bar{r^2Y_{2\mu}}}$ are expressed 
in terms of the double-stretched coordinates
$\bar q_i=\frac{\omega_i}{\omega_0} q_i$, $(q_i=x,y,z)$.
We remind that the  self-consistent residual interaction does 
not affect the equilibrium deformation obtained from the 
minimisation procedure.

Using the transformation from $p_i,q_i$ variables to the 
quanta $a_k^+,a_k$ \cite{Zel} ,
all matrix elements are calculated analytically.
We solve the RPA equation of motion for the generalised coordinates
${\cal X}_{\lambda}$ and momenta ${\cal P}_{\lambda}$
\bea
[H_{{\rm RPA}},{\cal X}_{\lambda}]&=&
-i\omega_{\lambda}{\cal P}_{\lambda},\quad
[H_{{\rm RPA}},{\cal P}_{\lambda}]=
i\omega_{\lambda}{\cal X}_{\lambda}, \\  \nonumber
[{\cal X}_{\lambda},{\cal P}_{\lambda}]&=&
i\delta_{\lambda, \lambda^\prime }.
\eea
where $\omega_{\lambda}$ are the RPA eigen-frequencies in 
the rotating frame and the associated phonon operators are
$O_{\lambda}=({\cal X}_{\lambda}-i{\cal P}_{\lambda})/\sqrt{2}$.
Here
${\cal X}_{\lambda}=\sum_s X_s^{\lambda} {\hat{f}}_s$, 
${\cal P}_{\lambda}=i\sum_s P_s^{\lambda}{\hat{g}}_s$ are 
bilinear combinations of
the quanta $a_k^+,a_k$ such that 
$\langle [{\hat{f}}_s, {\hat{g}}_{s^{\prime}}]\rangle=V_s\delta_{s,s^{\prime}}$ where
quantities $V_s$ are proportional to different combinations of $\sum_i$ 
$(i=x$,$+$,$-)$. Further, $\langle ... \rangle$ means the averaging over mean field states.
Since the mean field violates the rotational invariance, among the
RPA eigen-frequencies there exist two spurious solutions. 
One solution with zero frequency is associated with the rotation 
around the $x$-axes, since $[H,L_x]=0$. The other "spurious" solution 
at $\omega\equiv \Omega$ 
corresponds to a collective rotation, since 
$[H,L_{\pm}]=[H,L_y{\pm}iL_z]=\mp\Omega L_{\pm}$ \cite{m77}.
The Hamiltonian Eq.(\ref{hrpa}) possesses the signature symmetry, i.e.
$[R_x,H_{\rm RPA}]=0$ ($R_x=e^{-i\pi {\hat{L}}_x}$), such that it
decomposes into positive and negative signature terms
\beq
H_{\rm RPA}=H(+)+H(-) 
\eeq
which can be separately diagonalized  \cite{m77,JM,KN}. The negative signature
Hamiltonian contains the rotational mode and  
the vibrational mode describing
the wobbling motion \cite{JM,m79}. 
We focus on the positive signature Hamiltonian.  
It contains the zero-frequency mode 
defined by 
\beq
\label{an}
[H(+),\phi_x]=\frac{-iL_x}{{\cal J}_{TV}}, \quad
[\phi_x, L_x]=i
\eeq
and allows one to determine the
Thouless-Valatin moment of inertia ${\cal J}_{TV}$  \cite{MW}.
Here, the angular momentum operator $L_x=\sum_s l_s^x{\hat{f}}_s$
and the canonically conjugated angle $\phi_x=i\sum_s\phi_s^x{\hat{g}}_s$
are expressed via ${\hat{f}}_s$ and ${\hat{g}}_s$ which obey the condition
$R_x{\hat{d}}_sR_x^{-1}=\hat{d}_s$ $(\hat{d}_s={\hat{f}}_s$ or ${\hat{g}}_s)$. 
Solving Eqs.(\ref{an}) for the Hamiltonian $H(+)$,
\beq
H(+)=\sum_{k=x,+,-}\hbar\omega_k(a_k^{\dagger}a_k+1/2)-
\frac{\kappa}2(Q_0^2+Q_1^{(+)2}+Q_2^{(+)2})
\eeq 
where
\bea
&&Q_0 = \sqrt{\frac{5}{16\pi}}(2{\bar z}^2-{\bar x}^2-
{\bar y}^2) = \sqrt{\frac{5}{16\pi}}
\sum_s q_s^0{\hat{f}}_s\\
&&Q_1^{(+)} = \sqrt{\frac{15}{4\pi}}{\bar y}{\bar z} = \sqrt{\frac{15}{4\pi}}
i\sum_s q_s^1{\hat{g}}_s\\
&&Q_2^{(+)} = \sqrt{\frac{15}{16\pi}}({\bar x}^2-{\bar y}^2) =
\sqrt{\frac{15}{16\pi }}\sum_s q_s^2{\hat{f}}_s
\eea
\vskip 1cm
we obtain the expression for the Thouless-Valatin moment of inertia
\beq
\label{tv}
{\cal J}_{TV}={\cal J_{I}}+\frac{[2S_{x0}S_{x2}S_{02}-S_{x0}^2(S_{22}-\frac{1}{\kappa_2})
-S_{x2}^2(S_{00}-\frac{1}{\kappa_0})]}{[(S_{00}-\frac{1}{\kappa_0})
(S_{22}-\frac{1}{\kappa_2})-S_{02}^2]}
\eeq
Here, the term ${\cal J_{I}}$ corresponds to the Inglis moment of 
inertia 
\beq
\label{ing}
{\cal J_{I}}=\sum_s \frac{(l_s^x)^2V_s}{E_s}.
\eeq
The second term in Eq.(\ref{tv}) is a 
contribution of the quadrupole residual interaction in the cranking
model. In the cranking harmonic oscillator it consists of the terms which
have the following structure
\beq
S_{xm}=\sum_s \frac{l_s^xq_s^mV_s}{E_s},\quad S_{nm}=\sum_s
\frac{q_s^nq_s^mV_s}{E_s}, \quad n,m=0,2
\eeq
where $E_s$ are the energies of particle-hole excitations: 
$E_1=2\hbar \omega_+$, $E_2=2\hbar \omega_-$, $E_3=2\hbar \omega_x$, 
$E_4=\hbar\omega_+ + \hbar \omega_-$
and $E_5=\hbar \omega_+ - \hbar \omega_-$.
We also introduced the following notations:
$\kappa_0=\frac{5}{16\pi}\kappa$ and $\kappa_2=\frac{15}{16\pi}\kappa$.

The above results are the starting point for our numerical analysis.
To take into account shell effects we consider two systems with number 
of particles  $A=20, 64 \quad (N=Z)$. 
For $\hbar \Omega=0$ MeV the global minimum occurs for a prolate 
shape and for a near oblate triaxial shape for  $A=20$ and $64$, 
respectively \cite{HN}. If we trace the configurations 
which characterise the ground states, 
with increasing rotational frequency  both systems 
become oblate. At this point the moment of inertia vanishes, since there is 
no a kinetic energy associated with such a rotation.

\begin{figure}
\noindent
\centering
\includegraphics[width=7.0cm,angle=270]{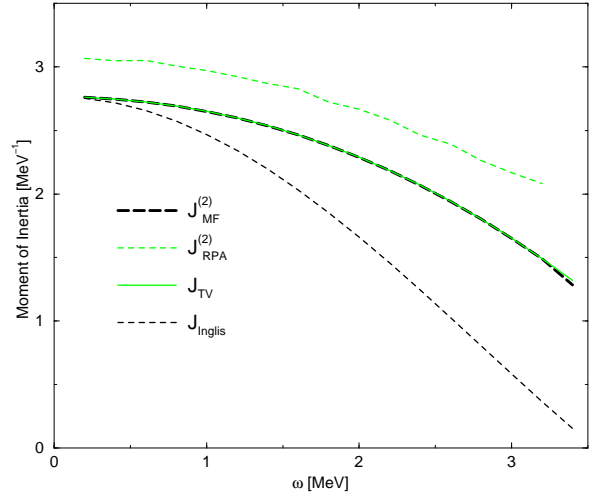}\\
\caption{Moments of inertia for $N=Z=10$ system as a function
of the rotational frequency $\omega \equiv \Omega$. 
The definitions of different moments of inertia are given in the text.
}
\label{fig2}
\end{figure}

\begin{figure}
\noindent
\centering
\includegraphics[width=7.0cm,angle=270]{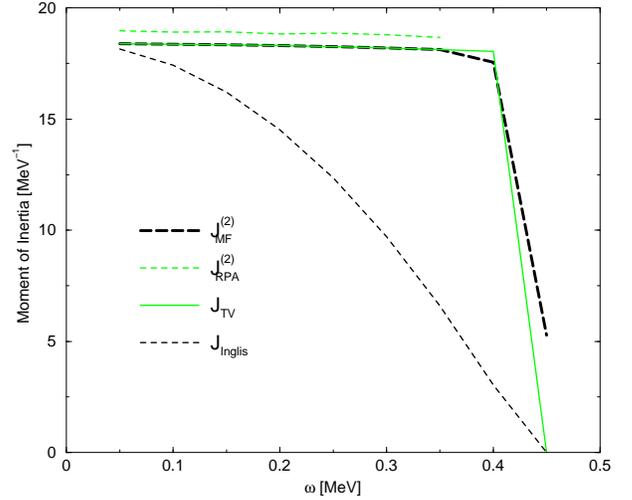}\\
\caption{As in Fig.\ref{fig2} for $N=Z=32$
system
}
\label{fig3}
\end{figure}

In order to compare various moments of inertia, i.e.
the  Thouless-Valatin, Eq.(\ref{tv}),
the Inglis, Eq.(\ref{ing}), and 
${\cal J}^{(2)}_{MF}=-d^2E_{MF}/d\Omega^2$ with
${\cal J}^{(2)}_{RPA}=-d^2E_{RPA}/d\Omega^2$ , 
we calculate the RPA correlation energy
$E^{RPA}_{\it corr}=\frac{1}{2}(\sum_{\lambda}\omega_{\lambda}-\sum_s E_s)$
which includes the positive and negative signature contributions. 
In Figs.\ref{fig2},\ref{fig3} 
the results of calculations for different moments of inertia are presented 
for $A=20$ and $64$, respectively. For our knowledge this is 
a first numerical demonstration of 
the equivalence between the dynamical moment of inertia 
${\cal J}_{MF}^{(2)}$
calculated in the mean field
approximation and  the Thouless-Valatin moment of inertia ${\cal J}_{TV}$ 
calculated in the RPA. For the both systems the Inglis moment of inertia 
${\cal J}_{I}$ is smaller than the ${\cal J}_{TV}$ and 
${\cal J}_{MF}^{(2)}$
and has a different rotational dependence. 
While the Inglis moment of inertia
characterises the collective properties of non-interacting fermions, 
the dynamical moment of inertia reflects the changes in the rotating 
self-consistent mean field due to an inter-nucleon interaction. 
As it was pointed out in \cite{m90}, the volume conservation
condition, used as a constraint in the mean field calculations, 
can be interpreted as a Hartree approximation applied to an
interaction that involves the sum of one-, two- etc forces. 

The sharp drop of all moments of inertia in Fig.2
is caused by the onset of the oblate shape where the collective rotation does
not exist. For $A=64$ the onset of the oblate deformation occurs at 
a smaller rotational frequency in contrast to the one for the system $A=20$. 

The dynamical moment of inertia ${\cal J}_{RPA}^{(2)}$ is larger
than the Thouless-Valatin moment of inertia. 
However, from our calculations it follows that the contribution of the
RPA ground state correlations decreases with an increase of the
number of particles.
The difference between the ${\cal J}_{RPA}^{(2)}$ and the 
${\cal J}_{TV}$ is
due to the following reason.
The Inglis moment of inertia  
is smaller than the Thouless-Valatin (or ${\cal J}_{MF}^{(2)}$) 
value, since the ${\cal J}_{TV}$ contains the effect of the 
residual particle-hole interaction.
On the other hand, the Thouless-Valatin moment of inertia manifests 
the rotational dependence of the residual interaction.
Thus, we may speculate that inclusion of the phonon 
interaction could help to reproduce 
the behaviour of the ${\cal J}_{RPA}^{(2)}$ which characterises 
the rotational dependence of the phonon-phonon interaction.

In summary, using the self-consistent cranking harmonic oscillator model, we
have numerically proved the equivalence of the dynamical moment of inertia 
calculated in the mean field approximation to the Thouless-Valatin moment of
inertia calculated in the RPA. Our result is a consequence of the 
self-consistent condition Eq.(\ref{cond}) which minimises 
the expectation value of
the mean field Hamiltonian, Eq.(\ref{mf}). 
This condition is equivalent to the  stability condition
of collective modes in the RPA \cite{T21}, i.e. 
$\omega_{\lambda}$ to be real and 
non-negative, and has been used to calculate 
different moments of inertia.  
The rotational dependence of the both  
dynamical moments of inertia, 
${\cal J}_{MF}^{(2)}$ and  ${\cal J}_{RPA}^{(2)}$, is  similar, however,
the ${\cal J}_{RPA}^{(2)}$ is larger than the 
${\cal J}_{MF}^{(2)}$ due to
the contribution of the ground state correlations.
This difference between the moments of inertia
is less important for heavy systems.

\vspace{.5cm}

{\it Acknowledgement:}  
This work was supported in part by the Heisenberg-Landau
program of the JINR.

\end{document}